# DFT Simulations of Inter-Graphene-Layer Coupling with Rotationally Misaligned hBN Tunnel Barriers in Graphene/hBN/Graphene Tunnel FETs

*Amithraj Valsaraj[1,a], Leonard F. Register [a], Emanuel Tutuc [a] and Sanjay K. Banerjee [a]*

[a] Microelectronics Research Center and Department of Electrical and Computer Engineering, The University of Texas at Austin, Austin, TX-78758, USA.

Van der Waal's heterostructures allow for novel devices such as two-dimensional-to-two-dimensional tunnel devices, exemplified by interlayer tunnel FETs. These devices employ channel/tunnel-barrier/channel geometries. However, during layer-by-layer exfoliation of these multi-layer materials, rotational misalignment is the norm and may substantially affect device characteristics. In this work, by using density functional theory methods, we consider a reduction in tunneling due to weakened coupling across the rotationally misaligned interface between the channel layers and the tunnel barrier. As a prototypical system, we simulate the effects of rotational misalignment of the tunnel barrier layer between aligned channel layers in a graphene/hBN/graphene system. We find that rotational misalignment between the channel layers and the tunnel barrier in this van der Waal's heterostructure can significantly reduce coupling between the channels by reducing, specifically, coupling across the interface between the channels and the tunnel barrier. This weakened coupling in graphene/hBN/graphene with hBN misalignment may be relevant to all such van der Waal's heterostructures.

## I. INTRODUCTION

The atomically thin two-dimensional (2D) materials translate to excellent electrostatic gate control at nanoscale channel length dimensions[1], near-ideal two dimensional carrier behavior[2,3], for conventional[4] and novel devices applications. Examples of the latter include field effect transistors with gate controlled resonant interlayer tunneling[5–8], which we here refer to as a class as interlayer tunnel FETs (ITFETs), and the proposed Bilayer-pseudospin FET (BiSFET)[9,10] based on interlayer exciton condensation. Experimental advances in stacking 2D atomic crystals have allowed the fabrication of van der Waal's heterostructures with precisely chosen sequence of atomic layers[11,12]. Stacking 2D materials in vertical van der Waal's heterostructures allows researchers to probe interesting physical phenomena such as interlayer tunneling between channel layers through a tunnel barrier, as demonstrated in a single layer (SL) graphene/hexagonal boron nitride (hBN)/SL-graphene heterostructure[13]. Aligning the band structures of the opposing graphene layers in momentum space then allows for resonant interlayer tunneling and associated negative differential resistance[7,8,14], which is the foundation for ITFETs.

Aligning the band structures in momentum space, of course, requires carefully aligning the crystallographic orientation of the two graphene layers. Achieving such alignment experimentally requires substantial care.[6–8,15] During layer-by-layer exfoliation of these van der Waal's materials, rotational misalignment is the norm, and even in such studies with aligned channel layers, the tunnel barrier has not been rotationally aligned. However, misalignment may affect device properties substantially. Prior theoretical studies have calculated the electronic structure of rotationally misaligned graphene bilayers using a continuum approximation[16] and a tight-binding method[17]. The opening of a band gap in the electronic structure of rotationally misaligned bilayer graphene has been demonstrated due to the formation of interlayer-bonded diamond nanodomains[18]. Variations in the electronic structure of graphene deposited on rotationally misaligned hBN even have been considered in a study of hydrogen desorption[19]. Density functional theory (DFT) calculations have been performed on short-period crystalline structures to derive effective Hamiltonians to describe the influence of moiré pattern superlattices on the electronic properties of graphene/graphene and graphene/hBN heterostructures[20]. However, prior theoretical studies of resonant interlayer tunneling have not considered the effect of rotational misalignment of tunnel barrier layers with respect to the channel layers on the tunneling current.[5,14,21]

---
[1] Author to whom correspondence should be addressed. Electronic mail: amithrajv@utexas.edu.

In this work, we consider the effects on interlayer current flow that rotational misalignment of the tunnel barrier with respect to rotationally aligned channel layers may have. Consider for the sake of illustration that the resonant tunneling currents $I$ of interest here could be approximated in the form $I \approx \eta e^{-\kappa t}$ at resonance, where $\eta$ and $\kappa$ are constants and $t$ is the tunnel barrier thickness. Misalignment between the graphene layers and the hBN barrier structures, in principle, could alter the evanescence coefficient $\kappa$. However, extrapolation of experimental results[7] consistent with the apparent value of $\kappa$ (corresponding to about an order of magnitude reduction in current per hBN layer) in a bilayer graphene/multilayer-hBN/bilayer graphene system to a few and even zero layers still leads to a prediction of weak interlayer tunneling even with the graphene bilayers aligned to each other (confirmed by resonant interlayer tunneling) but not to the hBN tunnel barrier. This limiting result cannot be explained through changes in evanescence alone. Therefore, we also must consider changes in the lead coefficient $\eta$ due to rotational misalignment, such as would be due to weakened coupling across the rotationally misaligned graphene-hBN interface. To consider these effects, we employ the *ab-initio* atomistic DFT approach. Direct calculation of current flow in this way, however, is impractical with actual ITFET device and current-flow geometries. However, DFT calculations allow us to track the effects of the hBN on the (nominal) graphene band structure as a measure of the effective coupling across the interface, as a function of rotation angle. Moreover, the inter-channel layer coupling that mediates this current flow also leads to degeneracy breaking in the band structure of the channel layers, so that knowledge of one provides information on the other. We consider a SL-graphene/hBN/SL-graphene systems, although with multi-layer (ML) hBN tunnel barriers in some simulations, largely to minimize the atom count in our DFT simulations with large in-plane unit cells in rotationally misaligned systems. However, the results may be relevant to all such van der Waal's heterostructures. We show that the rotational misalignment between the graphene and hBN layers can significantly degrade the interlayer-graphene-layer coupling, and that this reduction is consistent with the reduction in the coupling across the graphene-hBN interface. In contrast, we find that there is little apparent effect on the evanescence for the coupling within the hBN barrier. That is, our results are consistent with a reduction in the lead coefficient $\eta$ and not apparently with increases in the evanescence coefficient $\kappa$ in the above approximation for current. This interface-localized effect would be qualitatively consistent with the weak tunneling in the few hBN layer limit extrapolated from experimental results as noted above.

## II. METHODS

To translating knowledge of degeneracy breaking to information on current flow, consider that in the weak coupling limit—e.g., short of saturating the interlayer current flow to the lead-limited value—the tunneling current $I$ varies as

$$I = g_V \frac{4\pi e}{\hbar} \sum_{\mu\nu} |M_{\mu\nu}|^2 [f(E_\mu) - f(E_\nu)] \delta(E_\mu - E_\nu) \qquad (1)$$

within a first-order Bardeen transfer Hamiltonian approximation.[22,23] Here $M_{\mu\nu}$ represents the interlayer coupling, and is obtained from the overlap of the opposite channel layer wavefunctions across the interlayer tunnel barrier. Here, $g_V$ is the valley degeneracy of graphene and the summation in Eq. (1) runs over all states $\mu$, $\nu$ of the top and bottom graphene layers, respectively. However, we do not need numerical values for $I$ in this work; it is merely the quadratic dependence of interlayer coupling that is of interest. This interlayer coupling also produces a symmetric/antisymmetric splitting of the otherwise degenerate band structures in the opposite layers under flat band conditions, which is linearly proportional to $M_{\mu\nu}$ in the weak coupling limit. Thus, by directly calculating the energy splitting due to coupling through a given tunnel barrier, we also can infer the greater effect on interlayer tunneling currents, with the relative change in current being proportional to the square of the relative change in degeneracy breaking in the weak coupling limit.

For aligned and misaligned systems, and for SL and ML hBN barriers, DFT simulations were performed to calculate both the absolute band structures and the band splittings due to inter-graphene-layer coupling in the rotationally aligned and misaligned systems. However, first, the atomic structures were relaxed using the projector-augmented wave method with a plane-wave basis set as implemented in the Vienna *ab initio* simulation package(VASP)[24,25]. Graphene/hBN/graphene supercells were created using the Virtual NanoLab software[26]. The lattice constants of graphene and hBN are well matched (to within 1.6%), and the considered rotational misalignments between the hBN and the mutually aligned graphene layers are necessarily commensurate for our calculations. Therefore, lattice strain is minimal for the considered supercell structures. Commensurate rotation angles for the graphene/hBN interface are given by,[16]

$$\theta = \cos^{-1} \frac{3n^2 + 3n + 0.5}{3n^2 + 3n + 1}, \qquad (2)$$

where $n$ is a non-negative integer ($n=0,1,2,......$). The number of atoms in the supercell is given by $3 \times (6n^2 + 6n + 2)$. Figure 1(a) and (b) show the commensurately rotated graphene/SL-hBN/

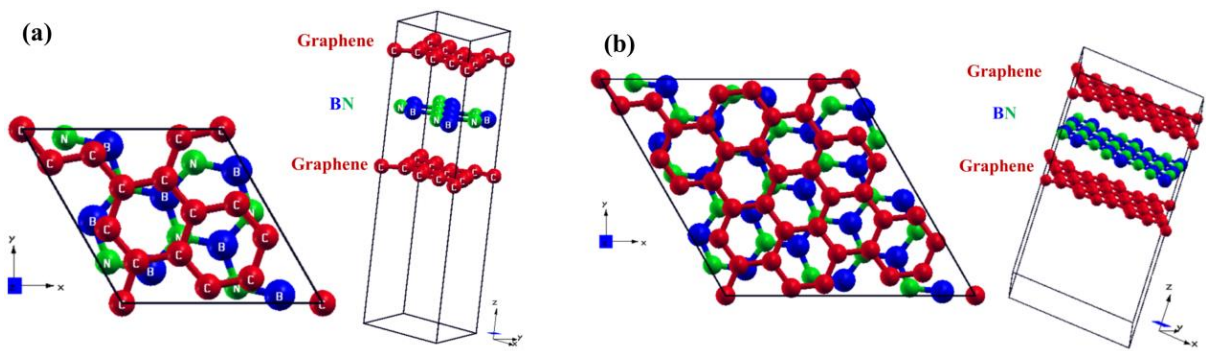

FIG. 1. Supercells of graphene/SL-hBN/graphene with hBN layers commensurately rotated with respect to mutually aligned graphene layers, at rotation angles of (a) 21.79° and (b) 13.17°. C atoms are colored red, B atoms in blue and N atoms in green online.

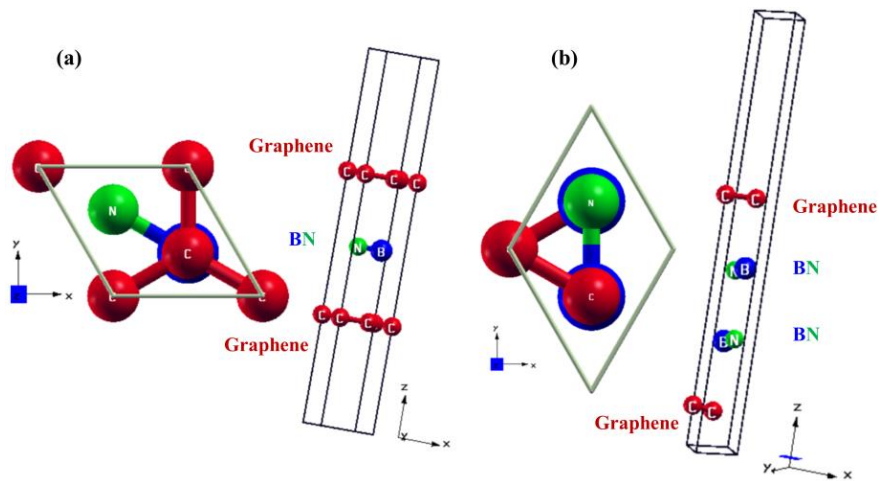

FIG. 2. Supercells of aligned graphene/hBN/graphene system with (a) SL-hBN and (b) bilayer hBN. C atoms of graphene are placed on top of B atoms of the hBN layer in the energetically most stable configuration.

graphene supercells with rotation angles, 21.79° and 13.17°, respectively corresponding to two smallest supercells ($n = 1,2$ from Eq. (2)) and visualized here using XCrySDen[27]. Hexagonal supercells with varying number of aligned hBN layers are shown in Figure 2. In the case of the aligned system, the C atoms of graphene are placed on top of B atoms of the hBN layer, which is the energetically most stable configuration[28]. With approximately 200 atoms per supercell, use of more computationally intensive hybrid functionals or GW methods is not practical. The local density approximation (LDA)[29] was employed for the exchange-correlation (XC) functional as LDA can model materials well with the same orbital character at conduction and valence band edges. In previous work on 2D materials on high-$k$ dielectrics, we had used LDA, generalized gradient approximation (GGA)[30] and hybrid HSE06[31] for XC functional in simulations of heterostructures with no qualitative differences in the results obtained[32,33]. Here, we also have re-checked some of the key results using the GGA functional. Van der Waal's non-local dispersive forces were modeled using the DFT-D2 scheme wherein a semi-empirical correction is added to the conventional Kohn-Sham DFT theory[34]. A k-mesh grid of 7×7×3 for the sampling of the first Brillouin zone of the supercell was selected according to Monkhorst-Pack type meshes with the origin being at the Γ point for all calculations except the band structure calculation (where a fine k-space resolution along major axes was sampled). Atomistic relaxations of supercells were allowed to converge when the Hellmann-Feynman forces on the atoms were less than 0.001 eV/Å. After relaxation, inversion symmetry was enforced on the crystal structure so that the induced degeneracy splitting is only due to coupling between the graphene layers across the hBN layer(s) and not any level of incomplete relaxation. The interlayer degeneracy splitting calculated in this way is only weakly **k**-dependent within a few hundred meV of the Dirac point. For specificity, we calculated the splitting at ≈100 meV above the Dirac (**K**) point, avoiding the intersecting band structures about the Dirac point, as illustrated in Figure 3.

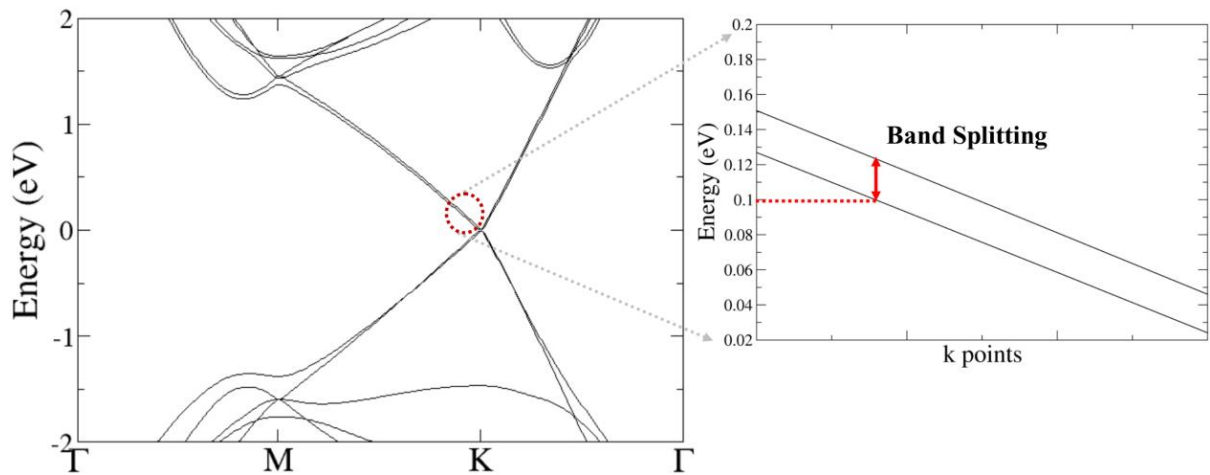

FIG. 3. An illustration of the method used to calculate the interlayer coupling parameter ($M_{\mu\nu}$) for the sample band structure of rotationally misaligned ($\theta = 21.79°$) graphene/SL-hBN/graphene after relaxation. The (roughly constant) band splitting is measured ≈100 meV above the Dirac (**K**) point.

**III RESULTS**

The interlayer energy splittings with hBN layer number for the fully-aligned graphene/hBN/graphene system calculated by the above outlined procedure were 69.1, 20.4, 2.2, and 0.49 meV for one to four layers of hBN, respectively (which is reasonably in line with the known zero hBN layer energy splitting limit of 370 meV represented by a Bernal stacked graphene bilayer[35]). For one, two or even more layers, such coupling essentially would short the graphene layers together in the large area systems, in contrast to the experimental results for rotationally misaligned systems[7,8].

To investigate this discrepancy between theory and experimental results, we first simulated channel/tunnel-barrier/channel heterostructures with a rotationally misaligned tunnel barrier layer between aligned channel layers. For SL-graphene/SL-hBN/SL-graphene system with the hBN layer misaligned by commensurate rotation with mutually aligned graphene layers, such as for a resonant tunneling device), Figure 4 and Table 1 exhibit a substantial reduction in the energy splitting with layer misalignment. As also shown in Table I, such a reduction in interlayer coupling due to this interface-localized effect would drop the estimated interlayer current per unit area by factors of approximately 9, 25 and 53 for the three twist angles of 21.79°, 13.17° and 9.43°, respectively, within the 1st order perturbation limits for energy splitting and tunneling current (Eq. (1)). Thus, despite the growth in primitive unit cell size by factors of 7, 19 and 37 respectively, the expected interlayer current *per primitive unit cell* for these three rotation angles would remain roughly constant and actually decrease somewhat to 0.80, 0.76 and 0.70 times that for the aligned system. Although, we are not in the perturbational limit for the strongest energy splittings, these results nevertheless suggest a strong effect of rotational misalignment on current flow.

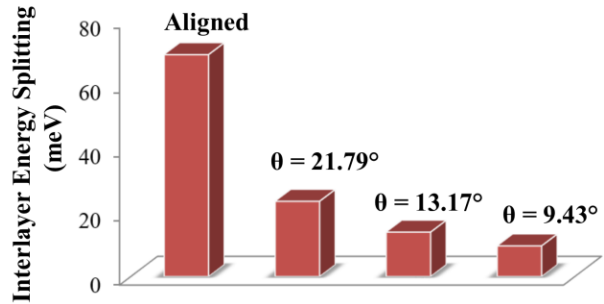

FIG. 4. Bar chart of interlayer energy splitting as function of twist angle ($\theta$) for graphene/SL-hBN/graphene heterostructures on a linear scale.

TABLE I: Variation of interlayer energy splitting as function of commensurate rotation angle for graphene/SL-hBN/graphene, along with number of atoms per unit cell and effect on current in the perturbative limit.

| Rotation Angle (degrees) | Degree of Band Splitting (meV) | No. of atoms in Supercell | Relative Current Drop (w.r.t. aligned case) |
|---|---|---|---|
| Aligned | 69.1 | 6 | -- |
| 21.79° | 23.4 | 42 | 8.72 |
| 13.17° | 13.8 | 114 | 25.1 |
| 9.43° | 9.51 | 222 | 52.8 |

We then considered SL-graphene/ML-hBN/SL-graphene heterostructures for graphene/hBN interfaces of fixed rotation angle $\theta = 21.79°$, corresponding to the smallest non-rotationally-aligned supercell from Eq. (2), and varied the number of hBN layers from 1 to 3. After relaxation of the supercell structures, the energy degeneracy splittings were computed and compared with the corresponding values for the aligned systems, and the effect on current in the perturbational limit was estimated, as shown in Table II. Although the reduction in energy degeneracy with increasing number of hBN layers is not entirely logarithmic for the aligned or misaligned systems (Figure 5)—perhaps not unexpectedly with potentially complex tunneling paths among atomic orbitals varying with the number of hBN layers within these atomistic systems—there is no evidence that increasing the number of layers would reduce the tunneling current more rapidly for a misaligned system. For this limited set of data, the effects of increasing tunnel barrier thickness are comparable to or, indeed, even weaker for the misaligned system. That is, to the extent that one may approximate interlayer current flow in the form $I \approx \eta e^{-\kappa t}$ for the sake of discussion, the results of these simulations, Tables I and II and Figures 4 and 5, are consistent with significant reductions in the lead coefficient $\eta$ and not with reduction in the evanescence coefficient $\kappa$ with increasing unit cell size.

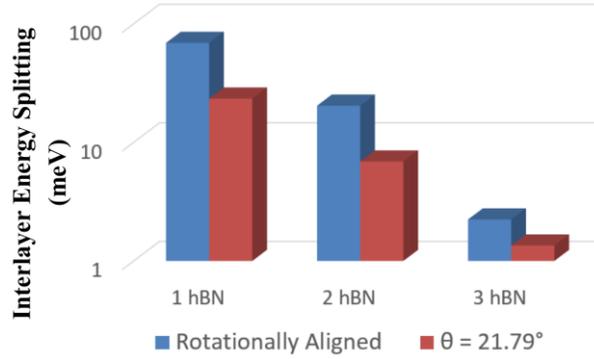

FIG. 5. Bar chart of interlayer energy splitting as function of tunnel barrier hBN layer number for rotationally aligned (blue) and twist angle $\theta = 21.79°$ (red) graphene/hBN/graphene heterostructures on a log scale.

TABLE II: Variation of interlayer energy splitting as function of layer number for aligned and rotationally misaligned graphene/hBN/graphene and effect on current in the perturbative limit.

| No. of hBN layers | Band Splitting for aligned system (meV) | Band Splitting for $\theta = 21.79°$ (meV) | Relative Current Drop (w.r.t. aligned case) |
|---|---|---|---|
| 1 | 69.1 | 23.4 | 8.72 |
| 2 | 20.4 | 6.90 | 8.74 |
| 3 | 2.24 | 1.35 | 2.76 |

In Figure 6, we also show the band-structure near the K point (nominal Dirac point) for both the rotationally aligned SL-graphene/three-layer-hBN/SL-graphene system and the rotationally misaligned ($\theta = 21.79°$) SL-graphene/three-layer-hBN/SL-graphene system. For the aligned system, in addition to the degeneracy breaking due to inter-graphene-layer coupling, a substantial effect on the band structure is observed near the K-point due to the interaction of the graphene layer with the hBN layers as predicted previously[28]. However, for the misaligned system, there is little apparent effect on band structure near Dirac point, again suggesting weakened coupling across the graphene-hBN interface in the rotationally misaligned case.

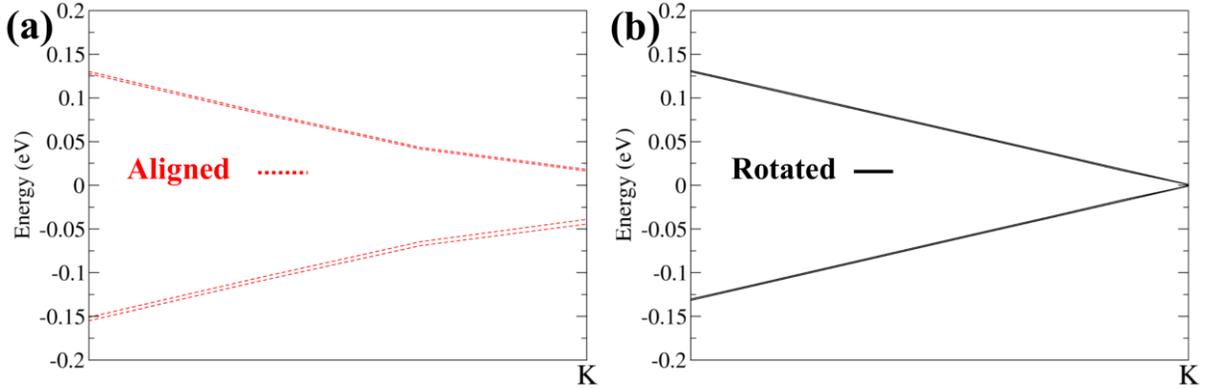

FIG. 6. A zoomed-in view of the band structure near the K point (nominal Dirac point) for (a) rotationally aligned SL-graphene/three-layer-hBN/SL-graphene system and (b) rotationally misaligned ($\theta=21.79°$) SL-graphene/three-layer-hBN/SL-graphene system.

For graphene/SL-hBN/graphene system with misaligned hBN layer between mutually aligned graphene layers, we repeated the simulations with GGA exchange-correlation functional for comparison with the LDA results. The same relaxed crystal structure was used, but separate self-consistent field and band structure simulations were performed using the GGA. Figure 7 shows the band structure and interlayer energy splitting near the Dirac point for graphene/SL-hBN/graphene with hBN layer rotated at an angle $\theta = 21.79°$, obtained using both the GGA and LDA functionals. As can be seen, the degree of band splitting matches closely. The interlayer energy splitting as function of twist angle for graphene/SL-hBN/graphene calculated using LDA and GGA have been listed in Table III for comparison. The energy splitting obtained using GGA functional is a few meV smaller than the corresponding LDA ones for all three cases.

TABLE III: Comparison of interlayer energy splitting obtained using LDA and GGA functionals, as a function of commensurate rotation angle for graphene/SL-hBN/graphene.

| Rotation Angle (degrees) | Band Splitting from LDA (meV) | Band Splitting from GGA (meV) |
|---|---|---|
| Aligned | 69.1 | 64.7 |
| 21.79° | 23.4 | 20.6 |
| 13.17° | 13.8 | 10.4 |

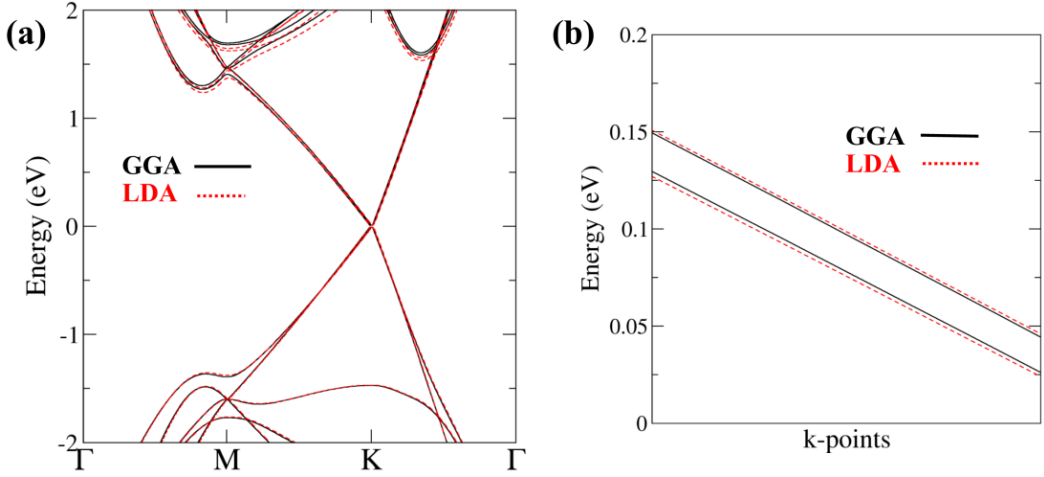

FIG. 7. (a) Band structure of SL-graphene/SL-hBN/SL-graphene with hBN layer rotated at $\theta = 21.79°$, obtained using GGA (black solid lines) and LDA (red dashed lines) exchange-correlation functionals. (b) The zoomed-in interlayer energy splitting near the Dirac point matches closely for both GGA and LDA.

Finally, we point out that as the rotation angle decreases, the interlayer coupling and the expected tunneling current does not tend to the value of aligned heterostructure in these calculations, but decreases instead as the unit cell size increases. In practice, for small rotation angles or, more generally, larger unit cell sizes beyond our capabilities to simulate, loss of coherence on unit cells dimensions or other causes would likely lead to saturation of the interlayer current density, which we expect is the experimentally observed condition. Our results only suggest that rotation leads to lower than otherwise expected interlayer tunneling currents, without addressing the limit of this reduction. Moreover, for sufficiently small rotation angles, the physical structure may self-align, at least locally,[36] leading back to a strong interlayer coupling and current flow.

## IV. CONCLUSION

In summary, our simulation results suggest that the interlayer tunneling current for graphene/hBN/graphene heterostructures will be affected strongly by the rotational alignment of the hBN interface with respect to the graphene layers. Indeed, for the three commensurate rotation angles considered in our study (21.79°, 13.17° and 9.43°), despite the growth in primitive unit cell size, the expected interlayer current per primitive unit cell for these three rotation angles remained roughly constant and actually decreased somewhat. At some point, as the rotation angle decreases or, more generally, the primitive unit cell size increases beyond our ability to simulate, this trend obviously must fail, perhaps due to loss of coherence on unit cells dimensions and, ultimately, perhaps local formation of commensurately aligned regions separated by non-aligned regions[36]. However, the essential behavior of interest, reduction of current flow with interlayer rotation should remain. Moreover, we did not see a greater effect on coupling and expected current reduction with increasing number of hBN layers in a misaligned system (21.79°) as compared to the aligned system. That is, to the extent that one may approximate interlayer current flow in the form $I \approx \eta e^{-\kappa t}$ for the sake of discussion, the results of these simulations are consistent with significant reductions in the lead coefficient $\eta$ and not with reduction in the evanescence coefficient $\kappa$ with increasing unit cell size. Accordingly, the reduction in expected current appears attributable to, specifically, the coupling across the interface between the channel layer and the tunnel barrier in this atomistic system, consistent with the weak tunneling current observed experimentally for such graphene/hBN/graphene systems[7,8]. Finally, it is perhaps reasonable to speculate that such effects of rotational misalignment also may be present in other van der Waal's heterostructures not directly simulated here.

## SUPPLEMENTARY MATERIAL

See supplementary material for the complete electronic structures of the studied rotationally misaligned graphene/hBN/graphene heterostructures.

## ACKNOWLEDGMENTS

This work was supported in part through the Southwest Academy of Nanoelectronics (SWAN) by the Nanoelectronic Research Inititiative (NRI), a Semiconductor Research Corporation (SRC) program sponsored by Nano Electronics Research Corporation (NERC) and National Institute of Standards and Technology (NIST). The authors acknowledge the Texas Advanced Computing Center (TACC) at The University of Texas at Austin for providing high performance computing (HPC) resources that have contributed to the research results reported within this paper.

## REFERENCES

[1] D. Jariwala, V.K. Sangwan, L.J. Lauhon, T.J. Marks, and M.C. Hersam, ACS Nano **8**, 1102 (2014).

[2] K.S. Novoselov, A.K. Geim, S.V. Morozov, D. Jiang, M.I. Katsnelson, I.V. Grigorieva, S.V. Dubonos, and A.A. Firsov, Nature **438**, 197 (2005).

[3] Y. Zhang, Y.-W. Tan, H.L. Stormer, and P. Kim, Nature **438**, 201 (2005).

[4] F. Schwierz, Nat. Nanotechnol. **5**, 487 (2010).

[5] P. Zhao, R.M. Feenstra, G. Gu, and D. Jena, IEEE Trans. Electron Devices **60**, 951 (2013).

[6] L. Britnell, R.V. Gorbachev, A.K. Geim, L.A. Ponomarenko, A. Mishchenko, M.T. Greenaway, T.M. Fromhold, K.S. Novoselov, and L. Eaves, Nat. Commun. **4**, 1794 (2013).

[7] B. Fallahazad, K. Lee, S. Kang, J. Xue, S. Larentis, C. Corbet, K. Kim, H.C.P. Movva, T. Taniguchi, K. Watanabe, L.F. Register, S.K. Banerjee, and E. Tutuc, Nano Lett. **15**, 428 (2015).

[8] S. Kang, B. Fallahazad, K. Lee, H. Movva, K. Kim, C.M. Corbet, T. Taniguchi, K. Watanabe, L. Colombo, L.F. Register, E. Tutuc, and S.K. Banerjee, IEEE Electron Device Lett. **36**, 405 (2015).

[9] S.K. Banerjee, L.F. Register, E. Tutuc, D. Reddy, and A.H. MacDonald, IEEE Electron Device Lett. **30**, 158 (2009).

[10] D. Basu, L.F. Register, D. Reddy, A.H. MacDonald, and S.K. Banerjee, Phys. Rev. B **82**, 75409 (2010).

[11] C.R. Dean, A.F. Young, I. Meric, C. Lee, L. Wang, S. Sorgenfrei, K. Watanabe, T. Taniguchi, P. Kim, K.L. Shepard, and J. Hone, Nat. Nanotechnol. **5**, 722 (2010).

[12] A.K. Geim and I.V. Grigorieva, Nature **499**, 419 (2013).

[13] L. Britnell, R.V. Gorbachev, R. Jalil, B.D. Belle, F. Schedin, A. Mishchenko, T. Georgiou, M.I. Katsnelson, L. Eaves, S.V. Morozov, N.M.R. Peres, J. Leist, A.K. Geim, K.S. Novoselov, and L.A. Ponomarenko, Science **335**, 947 (2012).

[14] R.M. Feenstra, D. Jena, and G. Gu, J. Appl. Phys. **111**, 43711 (2012).

[15] A. Mishchenko, J.S. Tu, Y. Cao, R.V. Gorbachev, J.R. Wallbank, M.T. Greenaway, V.E. Morozov, S.V. Morozov, M.J. Zhu, S.L. Wong, F. Withers, C.R. Woods, Y.-J. Kim, K. Watanabe, T. Taniguchi, E.E. Vdovin, O. Makarovsky, T.M. Fromhold, V.I. Fal'ko, A.K. Geim, L. Eaves, and K.S. Novoselov, Nat. Nanotechnol. **9**, 808 (2014).

[16] J.M.B. Lopes dos Santos, N.M.R. Peres, and A.H. Castro Neto, Phys. Rev. Lett. **99**, 256802 (2007).

[17] S. Shallcross, S. Sharma, E. Kandelaki, and O.A. Pankratov, Phys. Rev. B **81**, 165105 (2010).

[18] A.R. Muniz and D. Maroudas, Phys. Rev. B **86**, 75404 (2012).

[19] D.G. Kvashnin, S. Bellucci, and L.A. Chernozatonskii, Phys. Chem. Chem. Phys. **17**, 4354 (2015).

[20] J. Jung, A. Raoux, Z. Qiao, and A.H. MacDonald, Phys. Rev. B **89**, 205414 (2014).

[21] D. Reddy, L.F. Register, and S.K. Banerjee, in *Device Res. Conf. DRC 2012 70th Annu.* (2012), pp. 73–74.

[22] J. Bardeen, Phys. Rev. Lett. **6**, 57 (1961).

[23] S.C. de la Barrera, Q. Gao, and R.M. Feenstra, J. Vac. Sci. Technol. B **32**, 04E101 (2014).

[24] G. Kresse and J. Furthmüller, Phys. Rev. B **54**, 11169 (1996).

[25] G. Kresse and J. Furthmüller, Comput. Mater. Sci. **6**, 15 (1996).

[26] Virtual NanoLab version 2015.1, QuantumWise A/S (www.quantumwise.com).

[27] A. Kokalj, Comput. Mater. Sci. **28**, 155 (2003).

[28] G. Giovannetti, P.A. Khomyakov, G. Brocks, P.J. Kelly, and J. van den Brink, Phys. Rev. B **76**, 73103 (2007).

[29] J.P. Perdew and Y. Wang, Phys. Rev. B **45**, 13244 (1992).

[30] J.P. Perdew, K. Burke, and M. Ernzerhof, Phys. Rev. Lett. **77**, 3865 (1996).

[31] A.V. Krukau, O.A. Vydrov, A.F. Izmaylov, and G.E. Scuseria, J. Chem. Phys. **125**, 224106 (2006).

[32] A. Rai, A. Valsaraj, H.C.P. Movva, A. Roy, R. Ghosh, S. Sonde, S. Kang, J. Chang, T. Trivedi, R. Dey, S. Guchhait, S. Larentis, L.F. Register, E. Tutuc, and S.K. Banerjee, Nano Lett. **15**, 4329 (2015).

[33] A. Valsaraj, J. Chang, A. Rai, L.F. Register, and S.K. Banerjee, 2D Mater. **2**, 45009 (2015).

[34] S. Grimme, J. Antony, S. Ehrlich, and H. Krieg, J. Chem. Phys. **132**, 154104 (2010).

[35] K.F. Mak, M.Y. Sfeir, J.A. Misewich, and T.F. Heinz, Proc. Natl. Acad. Sci. **107**, 14999 (2010).

[36] C.R. Woods, L. Britnell, A. Eckmann, R.S. Ma, J.C. Lu, H.M. Guo, X. Lin, G.L. Yu, Y. Cao, R.V. Gorbachev, A.V. Kretinin, J. Park, L.A. Ponomarenko, M.I. Katsnelson, Y.N. Gornostyrev, K. Watanabe, T. Taniguchi, C. Casiraghi, H.-J. Gao, A.K. Geim, and K.S. Novoselov, Nat. Phys. **10**, 451 (2014).